\documentclass[aps,pre,preprint,groupedaddress,showpacs]{revtex4}

\usepackage{graphicx}
\begin{document}


\title{Approaches to Network Classification}


\author{Vladimir Gudkov }
\email[]{gudkov@sc.edu}
\affiliation{Department of Physics and Astronomy
\\ University of South Carolina \\
Columbia, SC 29208 }
\author{Joseph E. Johnson}
\email[]{jjohnson@sc.edu}
 \affiliation{Department of Physics and
Astronomy
\\ University of South Carolina \\
Columbia, SC 29208 }
\author{Shmuel Nussinov}
\email[]{nussinov@ccsg.tau.ac.il}
 \affiliation{Tel-Aviv University \\
 School of Physics and Astronomy \\ 
Tel-Aviv, Israel \\
 and  \\ 
 Department of Physics and Astronomy\\
  University of South Carolina \\
Columbia, SC 29208 }


\date{\today}

\begin{abstract}
We introduce a novel approach to description of networks/graphs. It is based on an analogue physical model which is dynamically evolved. This evolution depends on the connectivity matrix and readily brings out many qualitative features of the graph. 
\end{abstract}



\maketitle


\section{Introduction}
A graph or network consists of n vertices/nodes $V_i$ with edges (communication lines) connecting them.  It can be described by an $n \times  n$ connectivity matrix $C$ where

 \hskip1cm   $ C_{ij} = C_{ji} =$ number of edges connecting $V_i$ and $V_j$.
 
 Even when we allow $ C_{ij}$ to be only 0 or 1 - for (dis)connected $V_i V_j$, the number of $C$ matrices $2^{n(n-1)/2}$  is huge already for moderate $n$.

If two matrices differ only by the labelling of the vertices - i.e. by a similarity transformation $C^{\prime} =U^{-1}CU$ with $U$ $(U^{-1}=U^{\dag})$ effecting the permutation of rows (columns) of $C$ - then $C$ and $C^{\prime}$ represent the same graph.

Since there are n! such permutations the problem of deciding whether the two connectivity matrices correspond to the same graph is believed to be of a high degree of difficulty.  It is equally hard find intrinsic {relabelling invariant} features, of graphs which characterize {\em all} graphs. Even if not achieving this goal, such intrinsic features may be most valuable.  Thus the characteristic polynomial {or  eigen-value ($\lambda_1 \ldots \lambda_n$)} of the connectivity matrix encode many important graph theoretic features\cite{eigen}.

For most applications a complete characterization of graphs/networks is redundant.  We are often interested in the "Big picture" or gross features.  These include the answers to the following general questions about the graph/network: 

{$\bf Q _1$}: ``Are there some groups of vertices which are relatively strongly interconnected and more weakly connected to the rest of the ``external vertices'' ? ''

We will refer to these groups as ``clusters in graph.''  Clearly these differ from the graph theoretic ``cliques'' defined by requiring that each vertex in the clique be connected to all other vertices in the clique with no reference to the extent of external connections.  

$\bf Q_2$: ``Are there groups of vertices which are ``distant'' from each other in the sense that there are no (or few) ``short paths'' connecting them?''  (``Short paths'' are those with a small number of consecutive links.)

Ideally we would like to view a complex graph as a smaller set of ($k \ll n$) of ``super vertices'' each having a specific internal structure.  By connecting to other super vertices, these form a ``super graph'' at a higher level.  

The shear number of graphs seems to defy such a goal when all graphs are considered.  We believe however that actual communication, social, commercial, political etc networks  are essentially {\em not}  random.

The very history of their, often gradual, formation can result in a hierarchial clustering.  There is often a further tendency to enhance clustering.  If $V_i$ and $V_j$ are both strongly connected to $V_k$ then  $V_i$ and $V_j$ also frequently develop a direct connection.

Physical constraints such as the three dimensional space we live in and the essentially two dimensional surface of the earth or boards of printed circuits also play a crucial role along with the need to economize on the total length and usage of communication lines.

All the above tends to make ``clusters in graphs'' with relatively loose connections between them more likely so that two questions $Q_1$, $Q_2$ above can be answered in the affirmative.

The following analogue situation in biology is quite instructive.  An outstanding,  problem in post genomic biology is to predict the folding of proteins given their known amino acid sequence.
While natural ``native'' proteins almost instantaneously fold into their functional three dimensional form, artificially constructed, random, sequences do not.  It is believed that ``building blocks'' - and specific ``energy landscapes'' help guide the system to its correct folded form - in nature and in simulations.  This is reminiscent of the present problem where methods geared to specific ``Real Life'' networks with a presumed tendency for clustering are advantageous.

How can we efficiently search for such patterns?  

We can ask for the number of paths in the graph of length $s$ connecting a vertex $V_i$ to itself or $V_i$ to $V_j$.  By ``feeling out'' larger and larger region (as $s$ increases) we can tell if $V_i$ belongs inside a cluster and if $V_i$ and $V_j$ are distant in the sense described above.  We will elaborate on a simple approach for achieving this in Section II below.

Bringing vertices in a ``cluster'' into close spatial proximity can help in identifying these clusters.  This can be achieved in a dynamical approach in which we model the vertices $V_i$ by moveable point masses at $\vec{r}_i(t)$.  Attractive ``forces'' are postulated between any pair of points which are connected in the original graph.  Possible implementations of this general approach are discussed in Section III.

\section{The number of returning paths as a test for ``clustering in graphs''}

Imagine an actual physical model of the network where each edge is replaced by a $1\Omega$  resistor.  The electrical resistance between two nodes (or between two groups of vertices which are separately shorted) nicely models the ``distance'' between these nodes (or the two groups) as defined in $Q_2$ above.  The laws of adding resistances in series and in parallel imply that the resistance, like the ``distance'', increases the longer the paths on the graph connecting the two nodes are, and also decreases with the number of such connecting paths.
Instead of using this analog computation we can, by using powers of the connectivity matrix $C$,  trace out the evolution in $s$ steps of messages sent from each node to all its neighbors.  In fact the $i$, $j$ elements of $C^s$; $(C^s)_{ij}$ equals the number of paths  comprised of $s$ connected edges which start in $V_i$ and terminate in $V_j$.  In particular $(C^s)_{ii}$ is the number of paths returning to $V_i$ in $s$ steps.

When raised to a high power $C$, like any symmetric real matrix, simplifies considerably.  Let  $\lambda_1 \ldots \lambda_n$ be the $n$ real eigenvalues of $C$ in descending order and $\vec{V}_1 \ldots \vec{V}_n$  the corresponding orthonormal $n$ eigenvectors.  The columns   $\vec{C}^s$ of ${C}^s$ become all proportional to $\vec{V}_1$  with a factor representing the projection of $\vec{V}_1$   on the $i$-th column of $C$:
\begin{equation}
(\vec{C}^s)_i \propto (\vec{C}_i \cdot \vec{V}_i) \vec{V}_i
\end{equation}
and likewise for the rows.  Upon further multiplication by $C$, $C^s$ gets then multiplied by  $\lambda_i$.

For the special case when all vertices in $C$ have the same valency $v$ (i.e. each is connected to $v$ others)   $\lambda_i = v$ and 
\begin{equation}
V^+_i=\frac{1}{\sqrt{n}}(1,1, \ldots 1).
\end{equation}                                                             

While we seek some dilution of information such trivialization should be avoided.  Useful information can be obtained by looking at $(C^s)_{ij}$  at moderate values of $s$.  If i belongs in a rich heavily connected, ``cluster in graph'' then the initial rise of $(C^s)_{ij}$ :
\begin{equation}
(C^s)_{ii}\sim (v_{cl})^s \hskip1cm for\; i \in cluster 
\label{fastgr}
\end{equation} 
is higher than the initial rise of the same quantity when $V_i$ is a generic vertex located in a region of average ($\overline{v}$)   valency  so that:
\begin{equation}
(C^s)_{ii}\sim (\overline{v})^s
\end{equation} 
with  $\overline{v}\leq v_{cl}$.

To partially avoid the degeneration at high $s$, and gain more information from $C^s$ for large $s$, we tried adopting the following strategy.  Instead of $C^2$ we use 
\begin{equation}
\tilde{C}^2 = C^2 - diag\; C^2
\label{grrate}
\end{equation} 
Since the diagonal of $C^2$ counts all paths which come back to their origins in two steps, these paths are omitted in $\tilde{C}^2$.  Going one more step we consider $\tilde{C}^2\cdot C$.   By subtracting again its diagonal elements and defining 
\begin{equation}
\tilde{C}^3 = \tilde{C}^2\cdot C - diag\;\tilde{C}^2\cdot C
\end{equation} 
 we omit all paths which retrace in three steps, etc.  In general we define
\begin{equation}
\tilde{C}^{s+1} = \tilde{C}^s\cdot C - diag\;\tilde{C}^s\cdot C
\end{equation} 
And $(\tilde{C}^{s+1})_{ij}$   is the number of paths from $i$ to $j$ of length $s+1$ which have not formed at any prior stage a closed loop, and $(\tilde{C}^s\cdot C)_{ii}$  is the number of $i\rightarrow i$  such paths.(Such self avoiding walks are quite complex even for regular lattices (see, for exapmle recent papers\cite{ken,jaec}, and are tied to Ising models in the corresponding dimensions.  Exact enumerations of such SAW's yield solutions of the Ising models but not vice versa, parenthetically we note that if $\tilde{C}^{n-1}\cdot C$ has diagonal elements then the graph in question does have a hamiltonian circuit namely a closer path of length $n$ which visits each vertex exactly one time.)

While the latter number increases much more slowly than $(C^s)_{ii}$, it still ``runs-away'' as $s\rightarrow \infty$ , so that we need to ``re-normalize'' $\tilde{C}^s$   at each stage to have each $(\vec{\tilde{C}^s})_i$   column vector be of unit length.
A plot of $(\tilde{C}^{s-k}\cdot C)_{ii}$  as a function of $s$ could ideally help ``map out'' other clusters in the graph.  After ``exhausting'' all vertices in the putative initial cluster $C_1$ in which $i$ resided - which due to self-avoidance will take $n_1$ steps with $n_1$ the number of vertices in $C_1$ - we will wander off into a generic part of the graph. There the slower growth rate (\ref{grrate}) will take over.  If we can reach in $d_{i2}$ steps a second rich cluster $C_2$  we could after such number of steps start having again a fast growth rate (\ref{fastgr}).  This continues until $n_2$ steps later, $C_2$   is exhausted etc. etc.  

However the graph ``between the clusters'' is still a network.  This causes diffusive migration between two clusters with no sharp arrival times.  Also for appreciable $s$ several clusters may be reached at the same or similar number of steps.  These features tend to smooth out the changes of $(\tilde{C}^{s})_{ii}$.

\section{Dynamical Evolution highlighting network structure.}

A basic difficulty in discerning intrinsic graph / network structure is that the connectivity matrix depends on the labelling  of the vertices.  The following example clearly illustrates this.  Let us assume a large subset of vertices in our graph indeed divide naturally into fairly well-defined clusters  $C_1$ with $n_1$ vertices, $C_2$ with $n_2$ vertices etc up to $C_k$.  If we label our vertices in such a way that all vertices belonging in any one cluster are contiguous, the connectivity matrix will be ``Almost Block Diagonal''.
  
\begin{figure}[h]
\includegraphics{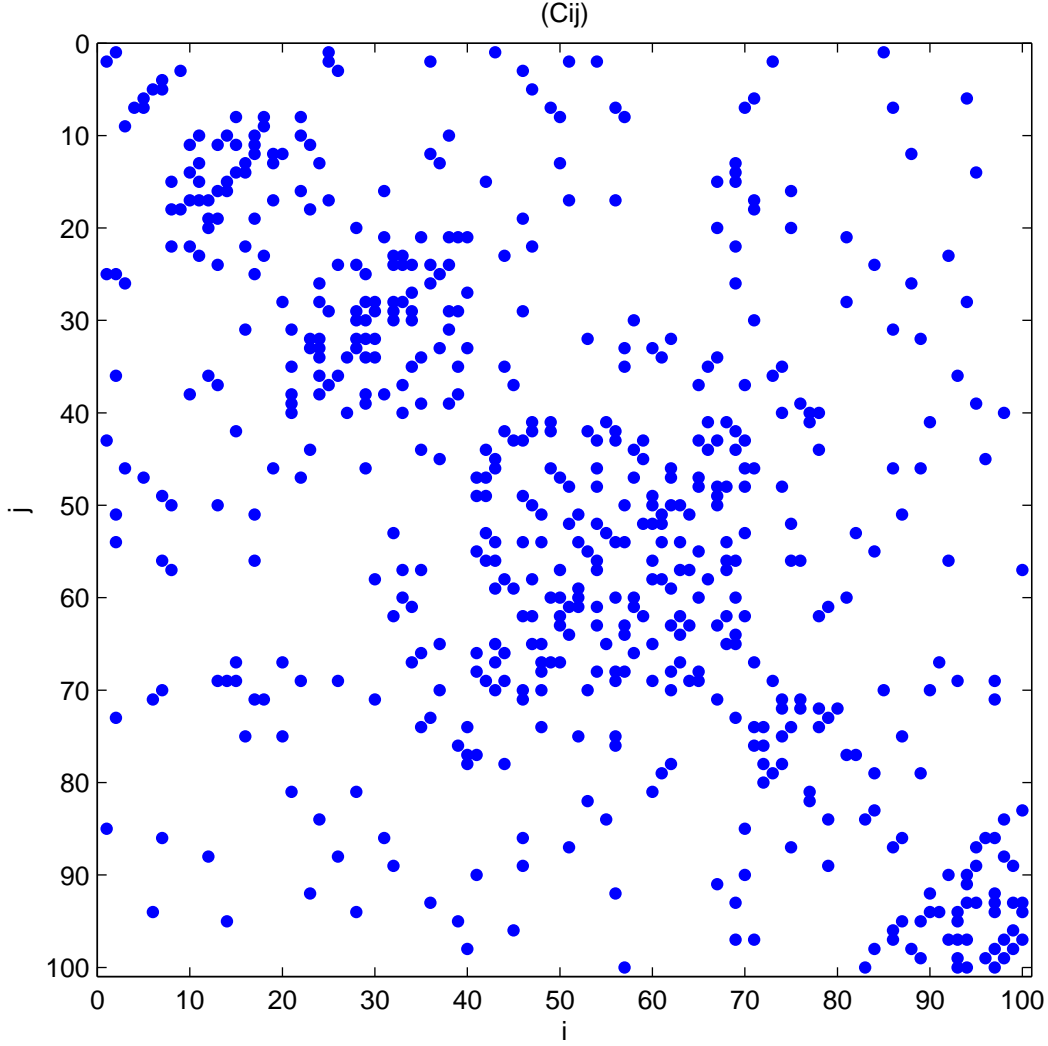}
\caption{Connectivity matrix with the average cluster valency 20\% and inter cluster connectivity valency 3\%. }
\label{fig:cblock}
\end{figure}

This is depicted in Fig.(\ref{fig:cblock}):  The $n_1 \times n_1$,  $n_2 \times n_2$ … sub matrices along the diagonal will then be connectivity matrices for the first, second, etc cluster.
By assumption these matrices have a relatively high proportion of non-vanishing elements.  The corresponding darker squares can thus be visually discerned relative to the background of the lighter more sparse remaining parts of the original $C$ matrix.

\begin{figure}[h]
\includegraphics{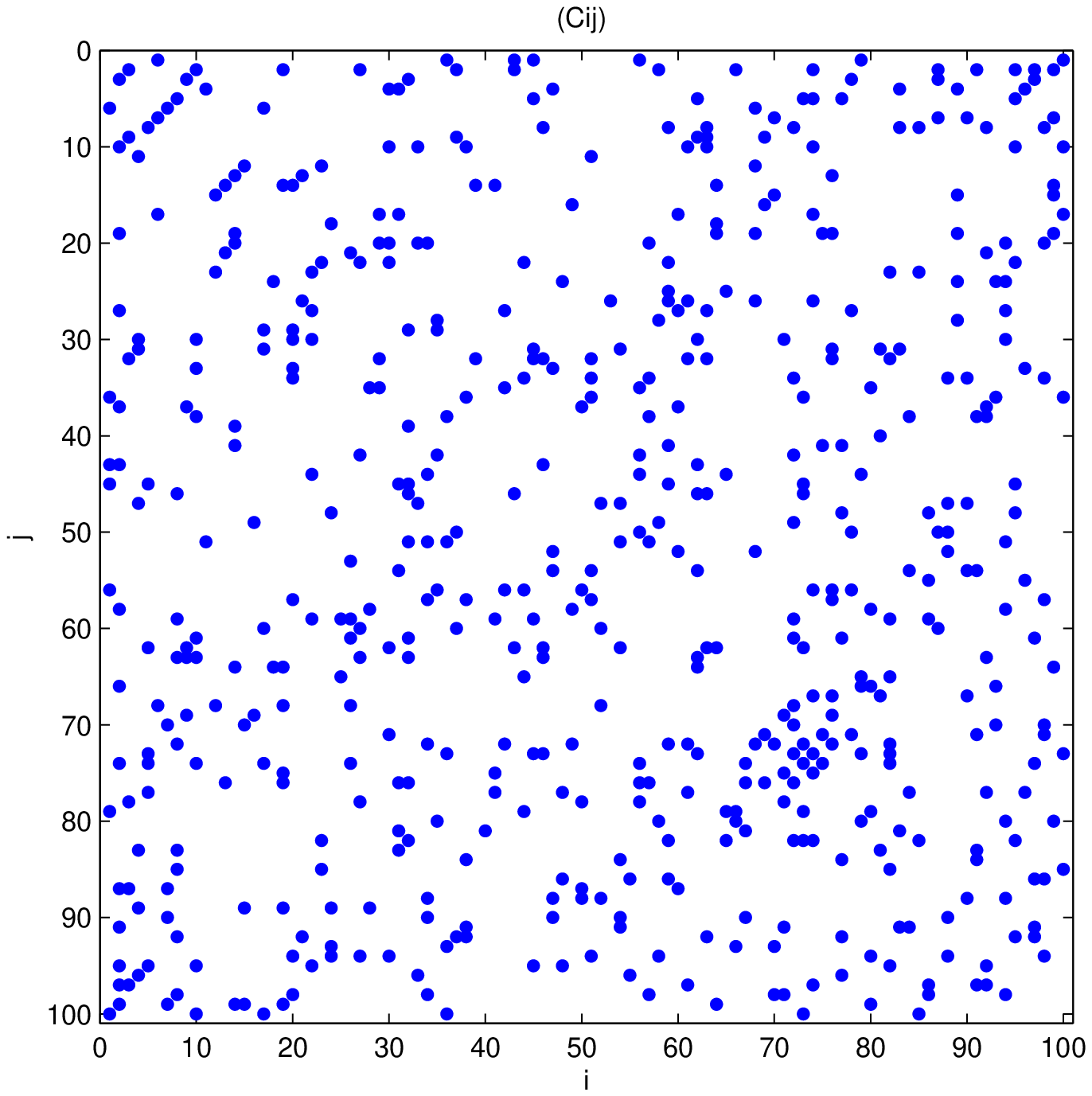}
\caption{Randomly reshuffled connectivity matrix $C$.}
\label{fig:b}
\end{figure}

This nice feature completely disappears after massive relabelling, i.e. massive joint reshufflings of columns and rows in the matrix $C$ (Fig.(\ref{fig:b})).  The whole matrix will then have a roughly constant average density of unit entries looking uniformly gray.  Our goal is essentially to reconstruct the original, convenient ``Almost Block Diagonal'' form which exhibits the clusters.  Its difficulty is exacerbated by the fact that the block diagonalization is only approximate and there are many non-vanishing entries outside the blocks.  Also we do not know a priori which size blocks and how many blocks do exist.
The representation of a graph by drawing it in two dimensions also introduces undesired arbitartrariness reflected in the choice of coordinates $(x_i, y_j)$ of the points representing the various vertices.  Two different drawings of the same graph may appear completely different and unrelated.

Such arbitrariness is particularly harmful when we try to implement the general idea described above and introduce attractive forces between any pair of points representing a pair of connected vertices.  The subsequent motion of the points may depend on their arbitrary initial placement.

To place the $n$ vertices in a completely symmetric and unbiased manner we need to go to $n-1$ dimensions.  The vertices (or the $n$ physical point masses modelling them in our approach) can be then put at the $n$ vertices of a symmetric simplex inscribed inside the unit sphere in $n-1$ dimensions.  Specific coordinates of the $n$ vertices can be constructed in a simple inductive process indicated in 
Appendix A.  All vertices are equidistant from the origin;  and specifically we chose:
\begin{equation}
\vec{r}^{\; 2}_i = 1                                                           
\end{equation}
using this,$(\sum \vec{r}_i)^2=0$   and the equality - due to symmetry - of all $\vec{r}_i \cdot \vec{r}_j$   for any $i \neq  j$ readily implies:
\begin{equation}
\label{scrr}
\vec{r}_i \cdot \vec{r}_j =-\frac{1}{n-1} \hskip1cm all  \hskip0.5cm  i \neq  j      \hskip0.5cm   i,j=1 \ldots n.                                              
\end{equation}
The distance between any pair of vertices of the simplex i.e. between any pair of the representative points at the outset of our proposed dynamical simulation is therefore:
\begin{equation}
|\vec{r}_i - \vec{r}_j|=\sqrt{\frac{2n}{n-1}}  \hskip1cm all \hskip0.5cm   i\neq j.                                         
\end{equation}
We next endow our system with some dynamics\cite{gold}.  We introduce a fictitious attractive force between points corresponding to vertices which are connected in the initial graph of interest.
Thus if $C_{ij}\neq 0$ we postulate
\begin{equation}
\vec{F}_{ij}(\vec{r}_i,\vec{r}_j)=\zeta_{ij}f(|\vec{r}_i - \vec{r}_j|)\frac{(\vec{r}_i - \vec{r}_j)}{|\vec{r}_i - \vec{r}_j|}.
\end{equation}                                                                         
To be the force attracting the point mass $i$ to the point mass $j$, in the direction of  $\vec{r}_i - \vec{r}_j$.  To retain the initial symmetry and avoid any biasing we take the same force law $f(r)$ for all pairs.  The specific shape of $f(r)$  will be tuned to optimize the gradual clustering.  In general $f(r)$ falls with distance and conversely, grows at short distances.

The only way information about the specific graph of interest is communicated to our dynamical n body system is via the overall strengths of the forces  $\zeta_{ij}$.  It vanishes if $C_{ij} = 0$.  For the generalizations considered later and also to better mimic real networks we allow any  $\zeta_{ij}\equiv  C_{ij} > 1$ so that it counts the number and ``quality'' of connections between $V_i$ and $V_j$. 

We next let our point move according to the standard newtonian dynamics:
\begin{equation}
m_i\frac{d^2\vec{r}_i}{dt^2}=\vec{F}_{i}=\sum_j\vec{F}_{ij}.
\end{equation} 
In order to avoid "overshoots" and oscillations we can add damping via viscous frictional forces:
\begin{equation}
m_i\frac{d^2\vec{r}_i}{dt^2}+\mu_i\frac{d\vec{r}_i}{dt} =\vec{F}_{i}.
\end{equation} 
Finally we can adopt the extreme $\mu_i\gg m_i$ so as to neglect inertial effects and have first order ``Aristotelian Dynamics'':  
\begin{equation}
\mu_i\frac{d\vec{r}_i}{dt} =\vec{F}_{i}.
\end{equation} 
The latter is readily discretized for time increments $\delta$:
\begin{equation}
\label{delr}
\vec{r}_i(t+\delta )=\vec{r}_i(t) +\frac{\delta}{\mu_i}\vec{F}_{i}(\vec{r}_i(t)) \hskip0.5cm  i=1 \ldots n, \hskip0.5cm  l\neq i \hskip0.5cm   l=1 \ldots n. 
\end{equation} 
To preserve the initial symmetry we take all initial mass ( and separately all initial viscosities) to be equal  $\mu_i=\mu $ , $m_i=m$.  Different masses (and / or viscosities) will arise at later stages when we treat super graphs with heavy vertices representing initial clusters.

The attractive central forces can be derived from a pair wise potential i.e.:  
\begin{equation}
f(r)=-\frac{d}{dr}U(r). 
\end{equation} 
And the overall potential energy is then:
\begin{equation}
\label{pot}
U(\vec{r}_1 \ldots \vec{r}_n)=\sum_{i>j}\zeta_{ij}U(|\vec{r}_i - \vec{r}_j|) 
\end{equation} 
 $U(r)$ is assumed to monotonically increase with decreasing $r$.  The possible equilibrium ``fixed points''of our dynamical system namely those for which 
\begin{equation}
\frac{d\vec{r}_i}{dt} =\vec{F}_{i}=0
\end{equation} 
 all $i$ are then stationary points of $U(\vec{r}_1 \ldots \vec{r}_n)$.
 
With only attractive forces or potentials present our $n$ point system eventually collapses towards the origin.  This is readily seen as the scaling
\begin{equation}
\vec{r}_i \rightarrow \lambda \vec{r}_i
\end{equation} 
with  $\lambda < 1$ will obviously decrease the $U(\vec{r}_1 \ldots \vec{r}_n)$  of equation (\ref{pot}) for any set of $\vec{r}_i$.

A joint collapse of all n points happening before the vertices belonging to ``clusters in the graph'' have separately concentrated in different regions compromises our goal of identifying the latter clusters.

To avoid the radial collapse we constrain $\vec{r}_i(t)$,  at all times to be on the unit sphere:
\begin{equation}
\label{constr}
|\vec{r}_i(t)|= constant = 1 \hskip0.5cm all  \hskip0.5cm t\geq 0.
\end{equation} 
To incorporate this we supplement eq.(\ref{delr}) by a length renormalization:
\begin{equation}
\vec{r}_i(t+\delta )\rightarrow \frac{\vec{r}_i(t+\delta )}{|\vec{r}_i(t+\delta )|}
\end{equation}
to be performed following the operation (\ref{delr}) at each step of our evolution.  The constraint (\ref{constr}) amounts to introducing normal (radial) reaction forces which cancel the radial components of any of the forces  $\vec{F}_{i}$, leaving us with only the tangential parts:
\begin{equation}
\vec{F}^T_i \equiv \vec{F}_i - (\vec{F}_i\cdot \vec{r}_i)\vec{r}_i
\end{equation}
The basic conjecture we make is the following:  
     
``After a sufficiently long time $T$ (or sufficiently many steps $s= T/\delta$) has elapsed    so that any point moved on average an appreciable distance away from its initial location  $|\vec{r}_i(T)-\vec{r}_i(0)|\geq a\approx 1$  geometrical clusters of points tend to form.  The points in each geometrical cluster correspond,  to a good approximation, to the original vertices in a ``cluster of the graph'' which these points represent.''

In the following we motivate this conjecture. 

We recall the definition of a cluster in the graph as a subset $C_l$ of $n_l$ vertices with a higher than average number of connections between them and than average number of connections with external vertices.  At $t=0$, the points representing any subset of $p$ vertices out of the $n$ vertices in the graph reside at the $p$ vertices of a $(p - 1)$ dimensional symmetric simplex.  All together there are $ \left(\begin{array}{c}
  n \\
  p \\
\end{array} \right)$ such ``faces'' of our original $n -1$ dimensional simplex.

To most clearly illustrate our point let us assume an ``ideal graph cluster'' so that in  first approximation we completely neglect those forces attracting members of the cluster (more precisely point masses representing vertices in the cluster) to ``outside'' points.
Had we also omitted the constraint (\ref{constr}) then the forces acting between the $n_l$ points of the cluster $C_l$ would initially and hence at all subsequent times, be restricted to the corresponding $n_l-1$ dimensional face.  Repeating the argument made originally for the full set of $n$ vertices, a collapse of these $n_l$ points into some point inside the $n_l$ simplex (i.e. on the $n_l-1$ dimensional face) is guaranteed.
With the constraint (\ref{constr}) enforced, the set of $n_l$ points will still collapse but now not to a point on the $n_l$ simplex but to a point on the ``spherical $n_l$ simplex'' which is the projection of the simplex on the unit sphere.  The point of common clustering need not be at the geometrical center of this spherical $n_l -1$ dimensional face.  However unless the cluster in question is very asymmetric in its internal connections, it may not be too far from it.

Let us next turn on the few forces pulling members of the cluster due to external vertices, i.e. points initially residing outside this face.  Such pulls may slightly shift the location of the clustering point away from the $n_l - 1$ dimensional spherical ``face''.   It is unlikely that it will disrupt completely the clustering of the vertices $V_i \in C_i$  belonging in the cluster.

We believe that the tendency to cluster will persist even in the more general case when the clusters are not so sharply defined. 

Let us focus on one particular vertex $V_i$ located at $t=0$ at $\vec{r}_i$ , one of the $n$ simplex vertices.  Among all the $\left(\begin{array}{c}
  n \\
  n_{l} \\
\end{array} \right)$  subsets of $n_l$ vertices, i.e. $n_l - 1$ dimensional faces, a subset of $\left(\begin{array}{c}
  n-1 \\
  n_l-1 \\
\end{array} \right)$  shares the specific $V_i$.  Stated differently,  $\left(\begin{array}{c}
  n-1 \\
  n_l-1 \\
\end{array} \right)$ different $n_l -1$ dimensional faces do intersect at the $V_i$ considered i.e. $n-1$ edges, $\frac{(n-1)(n-2)}{2}$  triangles,  $\frac{(n-1)(n-2)(n-3)}{3!}$  tetrahedrals and so on.  Furthermore each of the triangles includes two of the $n-1$ edges impinging at $V_i$, every tetrahedra contains three of these edges,… etc.

Let us next assume that among all such  $\left(\begin{array}{c}
  n-1 \\
  n_l-1 \\
\end{array} \right)$  simplexes there is a particular one which we denote by $S_l$ so that the point in question $\vec{r}_i$ , has a maximal number of forces acting in its direction (as compared with the number of forces acting on the direction of any one of the other simplexes).  This is the reflection in our dynamical model the fact that the vertex $V_i$ belongs in a cluster $C_l$ i.e. has more connections to $V_j \in C_l$ than to vertices in any other subset of $n_l$ vertices.

It is obvious that in the initially symmetric situation the point $\vec{r}_i$  will then start moving in the direction of that specific $n_l-1$ dimensional face since the force in its direction will be maximal.  The motion will not be {\em exactly} in this hyperplane as $V_i$ may have some external connections and consequently there will be forces on the mass point $\vec{r}_i$  in other directions.  However since the largest force component is along this direction the largest initial displacement  $\delta_1(\vec{r}_i)=\vec{r}_i(\delta )-\vec{r}_i(0)\; \propto \; \vec{F}_i$        also.  This motion will then be the first small step towards the formation of the physical cluster of the points representing $C_l$.

Now at $t=0$ all vertices start moving.  If all (or most) of the $n_l$ vertices on the simplex (face) in question share this same feature of $V_i$ then all (or most) of the $n_l$ points will tend to migrate away from the initial $n_l$ vertices of the simplex in question and move toward its interior. 
Once the representative points start to cluster on or near the corresponding $n_l -1$ dimensional spherical face the non-linear aspects of the many-body dynamical evolution come into play.  These will tend to enhance and accelerate the clustering in  several ways.

One such effect is simple.  As the group of points start to come closer together the average distances $|\vec{r}_i - \vec{r}_{i^\prime}|$  with $(V_i, V_{i\prime}) \in  C_l$ decrease.  By assumption the attractive forces between them become stronger.  This then accelerates the clustering of the points which started to cluster.

A presumably subtler effect is the more coherent pull on ``straggling vertices''.  These vertices belong to a strong cluster but due to ``Accidental'' connections to some different group of vertices start moving in a different direction. 

 The initial forces acting on any vertex have an angle of $60^o$ between any pair.    However because of our constraint of staying on the sphere we need to consider only the projection on the $n-2$ hyperplane tangent to the sphere at the vertex $V_i$, say, in question. After this projection the $n-1$ edges emanating from $V_i$ span the   $n-2$ dimensional hyperplane just in the same symmetric manner as the $n$ unit vectors $\vec{r}_i$  span the original $n-1$ dimensional $n$ simplex.  Hence at eq.(\ref{scrr}) the angle between members of any pair of the effective tangential forces is 
\begin{equation}
\cos{[\theta_{ij[projected]}]} = - \frac{1}{n-2}.
\end{equation}
Thus if $V_i$ was connected to  {\em all} the remaining $n-1$ vertices in the original graph the sum of all the (tangential!) forces acting on it would vanish.  In reality the valency of $V_i$, $v_i =$ total number of vertices directly connected to it is, much smaller than   $n-1$.  The almost orthogonal $v_i$ forces acting on it will thus tend to add in {\em quadrature}.  The same a-fortiori holds for the $v_{i{C_l}}$ forces directed to the face representing the cluster $C_l$.  ( $v_{i{C_l}}$ is a partial i-$C_l$ valency, namely the number of vertices in $C_l$ connected to $V_i$).

The initial force component along the $n_l -1$ dimensional face is then:
\begin{equation}
\label{forc}
\vec{F}_{i\{ i\in C_l\}}(t=0)=\sum_{j\in C_l}\vec{F}_{ij}(t=0)\;\propto \; [v_{i{C_l}}]^{1/2}
\end{equation}
Assume however that after some time most points corresponding to the putative cluster and in particular the $v_{i{C_l}}$ points in the cluster connected to $V_i$, have already bunched together on the surface of the sphere.  The various forces exerted by these $v_{i{C_l}}$ points on $V_i$ will now be almost {\em parallel} and instead of (\ref{forc}) we will have 
\begin{equation}
\label{forcf}
\vec{F}_{i\{ i\in C_l\}}(t>t_0)=\sum_{j\in C_l}\vec{F}_{ij}(t>t_0)\;\propto \; v_{i{C_l}}.
\end{equation}
The resulting force will be considerably enhanced if $v_{i{C_l}}>>1$.

If the vertices in the original graph had on average small overall valency then $v_{i{C_l}}$ could happen to be small - say $O(2-3)$.  The $\sqrt{v_{i{C_l}}}$  enhancement of (\ref{forcf}) relative to (\ref{forc}) would then be minimal.  Also $v_{i{C_l}}$ could be smaller than the number of connections that $V_i$ happens to have with points in some random face with $n_{c'1}=n_l-1$ dimensions.  The vertex $V_i$ will then ``wander off'' at $t=0$ in the direction of this face rather than that of the ``correct'' face corresponding to the cluster $C_l$.
We can avoid such situations and enhance the coherence effect discussed above by  replacing the original connectivity matrix by an appropriate power ($\tilde{C}^s$)  where the overall valencies (and in particular valencies pertaining to cluster) are (particularly) strongly enhanced.

Note that an ``error''due to an initial wandering off of $V_i$ in the direction of some random face which corresponds to no cluster in the graph, is corrected by the very clustering which is assumed to occur.  The other points in the ``random'' face will, by assumption, tend to migrate out of this face into other faces where these points can more efficiently cluster (physically).  Finding no nearby cluster on the wrong face the ``straying'' vertex $V_i$ in question is likely to be pulled back into the original cluster $C_l$ (or to another cluster which formed in the meantime and to which $V_i$ is more strongly connected).

Thus our dynamical evolution process can be construed not as just motion of $n$ points on the unit $n -1$ dimensional sphere, but rather we can view it as a competition between the putative different (physical!) clusters for additional members (points).  In this ongoing ``tug of war'' clusters with stronger internal connectivity are likely to ``win over'' farther members and form first.

Once the points corresponding to a cluster in the graph have ``bunched'' close together they become effectively one dynamical unit - a ``supervertex''.  Not only will all the points pull coherently external points but also the converse naturally holds: the clustered points will tend to respond coherently as one dynamical unit to an external force.  Thus assume that we try to ``pull away'' one member point.  Due to its close proximity to other members of the cluster the point in question will strongly pull on those connected to it.  The latter points in turn will pull on further points in the cluster etc and eventually the whole cluster will move in response to the external force.  Hence the compact clusters, once formed, will be stable with respect to ``stray'' external  ``tidal forces''.

The actual emergence of the physical clusters can be readily ascertained.  Once $|\vec{r}_i - \vec{r}_j |$  is smaller than a prescribed small number the pair of points are ``merged'' into one point, at $(\vec{r}_i + \vec{r}_j)/2$. Actually we need at this point to project again $(\vec{r}_i + \vec{r}_j)/2$ onto the sphere.  In further evolution the force acting on the merger point is the sum total of all the forces acting on $\vec{r}_i$  and $\vec{r}_j$ .  Also the resulting point should be endowed with twice the viscosity inertia $\mu_{i\cup j}=\mu_i+\mu_j$  and / or $m_{i\cup j}=m_i+m_j$.  This new, doubled up, point represents a new graph derived from the original by identifying $V_i$ and $V_j$.  It has $n-1$ vertices and its connectivity matrix has the same elements $C_{ll^{\prime}}$ when both $ll^{\prime}$ differ from either $i$ or $j$.  The new vertex $(V_{i\cup j} )$ is now connected to all the $ll^{\prime}$  vertices which were connected to either $i$ or $j$.

We can keep on merging using at each step the center of mass
\begin{equation}
\label{forc1}
\vec{r}_{cm(i,j)}\equiv \frac{m_i\vec{r}_i+m_j\vec{r}_j}{m_i+m_j}
\end{equation}
as the merge point.   Also we keep adding the masses and viscosities of the merged points and keep the connections to all vertices/points presently existing.

Ideally this process would yield, after a reasonable number of steps $s$, to $k$ ``supervertices'' corresponding to the $k$  blocks $\{n_{l_1}\times n_{l_1},\; n_{l_2}\times n_{l_2}, \ldots \; n_{l_k}\times n_{l_k}  \}$ in the properly ordered original $C$ matrix of Fig. 1.  The off-diagonal element $ll^{\prime}$ will be here the total number of ``$l$'' entries in the original matrix $C$ in the $n_{l}\times n_{l^\prime}$ rectangle at the ``intersection'' of the $C_l$, and $C_{l^\prime}$ blocks.  We could now repeat a similar dynamical procedure for the $k$ ``supervertices''.  This is in fact  what the above algorithm is doing anyway in a relatively smooth and continuous manner. 

Instead of merging pairs of closely aligned points, we can identify various physical clusters with some minimal number of points and merge those as above.  

The technicalities of how we actually merge the clustered points aside, the key question of whether the optimal desired clusters will form to start with, still remains.
The dynamical evolution described here forms clusters of all sizes - small ones with few members,  larger ones which may include all or parts of smaller clusters and the one big supercluster containing all $n$ vertices.  We have blocked, by constraining   at all times, the fastest route towards forming the overall cluster, namely via radial infall.  Clustering will eventually still occur at some point the unit sphere.

In structure formation in three dimensions, creation of small clusters requires the particles forming the cluster to travel for shorter distances than in the case of bigger clusters.  Due to the peculiar geometry of $n$ particles in $n-1$ dimensions described in appendix A, this intuition does {\em not} carry over.  Formation of any cluster requires roughly the {\em same} distance to be covered regardless of the size of the cluster.  Hence we are not guaranteed by essentially kinematic reasons that the smaller clusters will form first - en route to the bigger clusters - which is the desired scenario for our purposes here.

Careful tuning of the force of $f(r)$ helps achieve such a scenario.  For $f(r)\simeq c/r^{\alpha}$  with large $\alpha$ , small differences in the distances will have large effect, (note that for the gravitational force in $n-1$ dimensions, $\alpha =n-2$).  Too strong a rise for $r<r_{initial}$ and fall for $r > r_{initial}$ may however lead to accidental clustering of some small groups.  In particular it may diminish the effect of the corrective mechanism via the coherent pull of the elements of the cluster on straying elements described above. 

 An interesting alternative prevents complete clustering but allows formation of clusters with higher than average internal connectivity.  We introduce in addition to the above attractive force between vertices $V_i$ and $V_j$ with $C_{ij}\neq 0$, repulsive forces when $V_i$ and $V_j$ are {\em not} connected: 
\begin{equation}
G_{ij}=g(\vec{r}_{ij})\frac{(\vec{r}_i-\vec{r}_j)}{|\vec{r}_i+\vec{r}_j|}\equiv F_{ij}(r) \hskip0.5cm for \; C{ij}= 0.
\end{equation}
Again this can be derived from a repulsive $W(r)$ potential.

Since in general we have many more unconnected vertices in a graph with large $n$, the repulsion can be weak relative to the attraction.  Let us assume that the average valency is $v$.  If all $n$ points would physically cluster we have $O(n^2/2)$ repulsive interactions $W(a)$, with $a$ the size of the clustering region, and $O(nv/2)$ attractive interacter $V(a)$.  Thus it suffices to have 
\begin{equation}
W(a)\geq \frac{v}{n}V(a)
\end{equation}
in order to prevent forming complete clustering into one big supercluster.  (The constraint $|\vec{r}_i(t)|=1$   is still necessary to prevent vertices from being pushed to infinity!)

We note that as a putative new member is trying to join a cluster $C_l$, in which its valency $v_{i\{ C_1 \} }$  is higher than the average, we need to facilitate joining, to satisfy the following condition: 
\begin{equation}
\label{pot1}
W(a)\leq \frac{v_{i\{ C_1 \} }}{n_1}V(a).
\end{equation}
Since $v_{i\{ C_1 \} }\geq v$, and further $n_l << n$, we have a sizeable range of $W(a)/V(a)$ for which smaller clusters but not very large ones can first form.  By gradually phasing out the repulsive forces once the smaller clusters have formed,  we can proceed to forming bigger clusters etc.

The repulsive forces would tend to move to antipodal points on the sphere groups of points which are ``distant'' from each other in the graph theoretic sense of question 2 in the introduction.  Utilizing such forces would help identify such groups as well.

\section{Specific applications} 

\begin{figure}[h]
\includegraphics{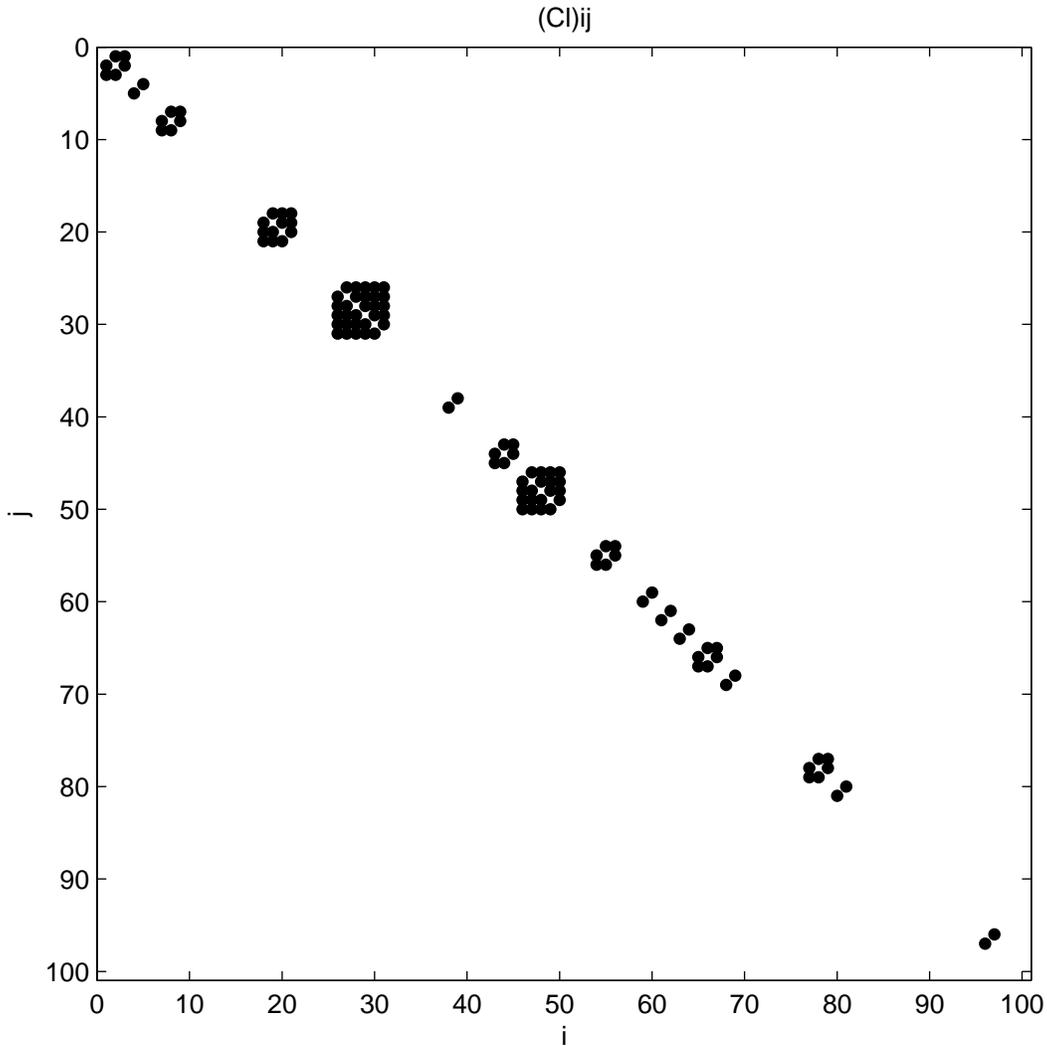}
\caption{Cluster connectivity matrix for reshuffled connectivity matrix $C$.}
\label{fig:rest}
\end{figure}

To demonstrate the power of our approach we applies it to the problem of  cluster identification in the 100-nodes network represented by the connectivity matrix $C$ of  Fig.(\ref{fig:cblock}). This matrix consists of seven clusters with randomly created internal connections with valency $20\%$. These clusters have been randomly interconnected with valency $3\%$.  To simulate a real-life situation of networks with unknown structure (topology) we randomly permutate the rows and columns of the matrix $C$ obtaining the reshuffled matrix $C^{\prime}$ shown in the Fig.(\ref{fig:b}). Next we apply our algorithm for clusters reconstruction using a combination of attractive and repulsive forces in $n-1=99$ dimensional space. The vertices of the 100-simplex were allowed to move under the influence of the forces on the 98-dimensional hyper-sphere in 99-dimensions. After a number of steps we analyzed the mutual distances between the vertices of the simplex and group neighbors which are close to each other into separate clusters. The new cluster-connectivity matrix is shown in Fig.(\ref{fig:rest}).
We see the seven ``big'' clusters of the matrix $C$ on a background of few small clusters due to the random (but still rather high) cluster inter connections.  The procedure identifies not only the cluster structure of networks but numerates and tabulates all the nodes in each cluster.  We father note that the distances in Fig. (\ref{fig:rest}) between the different clusters do - unlike the origin Fig. (\ref{fig:cblock}) - reflect the actual ``graph theory'' distance between them.      

\appendix
\section{} 
Some geometrical aspects of the $n$ simplex and its $p-1$ dimensional sub-simplex  faces are relevant to our dynamical evolution.  Most such features can be derived without utilizing any specific coordinate representation. 

The fundamental relation
\begin{equation}
\label{scrrA}
\vec{r}_i \cdot \vec{r}_j =-\frac{1}{n-1} \hskip1cm all  \hskip0.5cm  i \neq  j      \hskip0.5cm   i,j=1 \ldots n                                              
\end{equation}
was derived above by using $(\sum{\vec{r}_i})^2=0$  and symmetry.  It allowed us to deduce the length of any edge 
\begin{equation}
|\vec{r}_i - \vec{r}_j|=\sqrt{\frac{2n}{n-1}}  \hskip1cm all \hskip0.5cm   i\neq j                                         
\end{equation}
such edges can be viewed as 1-dim 2 point subsimplices.

We have also triangles, namely 3-simplices, forming 2-dim ``faces''/edges etc,  $\left(\begin{array}{c}
  n \\
  p \\
\end{array}\right)$   $p$-simplices etc.

Let $r_p$ denote the radius of the sphere circumscribing the $p$ simplex and $d_p$ the distance to its center from the origin (namely the center of the original $n$ simplex).  Clearly $d^2_p+r^2_p=1$.

Let $\vec{r}_{i_1} \ldots \vec{r}_{i_p}$  be the $p$ unit vectors of the $p$ simplex.  All the $i_p$ are different and there are  $\left(\begin{array}{c}
  n \\
  p \\
\end{array}\right)$ such possible subsets of the $n$ original $\vec{r}_i$.  The vector from the origin to the center of simplex is:  
\begin{equation}
\vec{d}_p = (\vec{r}_{i_1} + \vec{r}_{i_2}  + \ldots \vec{r}_{i_p} )/p                                              
\end{equation}
Hence using again (\ref{scrrA}) we find:
\begin{equation}
d_p = \sqrt{\vec{d}_p^{\; 2}} = \frac{1}{p}\sqrt{p-\frac{p(p-1)}{n-1}}=\sqrt{\frac{n-p}{(n-1)p}}.
\end{equation}
And
\begin{equation}
r_p = \sqrt{1-d_p^2} = \sqrt{\frac{n}{n-1}\cdot \frac{p-1}{p}},
\end{equation}
$r_p$   is the distance from vertex of the $p$ simplex to its center.  Except for very small $p$'s (representing ``tiny'' clusters) all $r_p$  are $O(1)$ so formation of such clusters would require the vertices to travel the same distance as in the formation of bigger clusters.

The actual angular separation between $\vec{r}_i$  in the $p$ simplex and $r_p$, the vector from the origin to its center is given by:
\begin{equation}
\theta_p=\arccos{(d_p)}=\frac{\pi}{2}-\arcsin{(r_p)}\approx \frac{\pi}{2} -  \sqrt{\frac{n-p}{(n-1)p}}.
\end{equation}
Two $p$ simplices can differ by just one, two $\ldots q$, $\ldots$   or $p-1$ points.  The distances $r_p^q$  between the centers of two neighboring $p$ simplices differing by $q$ vertices (and with $p-q$ common vertices) grow with $q$ for fixed $p$, as follows.

     The vector connecting the two centers is: 
\begin{equation}
\label{dA}
\vec{d}_p = \frac{1}{p}\left(\sum^q_{i=1}\vec{r}_i - \sum^q_{i=1}\vec{r}_{li}  \right).
\end{equation}
With the sets $\{ \vec{r}_i\} $,($\{ \vec{r}_{li}\} $) denoting the $q$ points in the first (second) $p$ simplices which are not shared by the two.  The common $\vec{r}_i$'s cancel in the difference,  and do not contribute to the distance $r_p^q$.  Using (\ref{dA}) and $\vec{r}_i \cdot \vec{r}_j =-1/(n-1)$  we find:
\begin{equation}
r^2_p = \sqrt{(\vec{d}_p^q)^2}=\frac{1}{p}\sqrt{2q+\frac{2q}{n-1}}.
\end{equation}
Hence the angle between  $\vec{r}_p^{\; (1)}$ the $\vec{r}_p^{\; (2)}$ vectors to the centers of the two simplices is given by: 
\begin{equation}
\theta^q_p = 2\arcsin{\left( \frac{r^q_p}{2d_p}\right)}=2\arcsin{\left[ \sqrt{\left( \frac{n}{n-p} \right)\frac{q}{2p}}\right]}.
\end{equation}
The last equation displays a nice feature.  There is a small angular distances between the centers of the (spherical) faces corresponding to two $p$ clusters which differ by a small fraction $q/p$ of their vertices.  The angular separation grows once $q\approx p \ll n$   to  $\theta_p^q \approx \pi /2$.

For our simulations we need an explicit representation of $\vec{r}_i$.  Assume we know the latter for the $n-1$ simplex (in $n-2$ dimension) denote them by $\vec{\vartheta}_1 \ldots \vec{\vartheta}_{n-1}$  with each $\vec{\vartheta}$ an $n-2$ vector with known components:
\begin{equation}
\vec{\vartheta}_j = \vec{\vartheta}_{j1}\hat{e}_1 + \ldots  \vec{\vartheta}_{j,n-2}\hat{e}_{n-2}
\end{equation}
with $\hat{e}_l$  the unit vector along the $l$-th axis.  When $n-1 \rightarrow  n$ we choose 
\begin{eqnarray}
  \vec{r}_n &=& \hat{e}_{n-1} \nonumber \\
   \vec{r}_i &=& \lambda_n \vec{\vartheta}_i - \frac{1}{n-1} \hat{e}_{n-1} \hskip0.5cm i=1 \ldots n-1.
\end{eqnarray}  
The normalizing factor  $\lambda_n= \sqrt{1-1/(n-1)^2}$ ensures $|\vec{r}_i|=1$, given that $|\vec{\vartheta}_i|=1$.  Thus, starting with a two simplex with $x^1_1=1; \; x^1_2=-1$, we inductively generate any $n$ simplex.

\begin{acknowledgments}
 S.Nussinov would like to thank Zohar nussinov for a crucial comment 
 regarding the advantage of going to higher dimensions to overcome 
 frustrations and alleviate constraints. He would like to dedicate this work to sir Isac Wolfson who donated the chair in theoretical physics at Tel-Aviv University on the occasion of his 80th birthday.
\end{acknowledgments}


\begin{thebibliography}{}



\bibitem {eigen} D. Cvetkovi\'{c}, P. Rowlinson and S. Simi\'{c}, ``Eigenspaces of graphs'' (Encyclopedia of mathematics and its applications, v. 66), Cambridge; New York : Cambridge University Press, 258p., 1997.
\bibitem {ken} T. Kennedy, J. Statist. Phys. {\bf 106}, 407 (2002).
\bibitem {jaec} A. Jaeckel and J. Dayantis, Macromol. Theory Simul. {\bf 10}, 461 (2001).
\bibitem{gold}An analogous physical system was used by Farhi, Goldstone and Gutmann and Sipser arXiv quant-ph/0001106 (2000). Their idea was to create a eave function of $n$ spins satisfying a set of Boolean logic logic requirments via adiabatic changing of the Hamiltonian.



\end{thebibliography}
\end{document}